# A novel feature in aluminum cluster photoionization spectra and possibility of electron pairing at T$\gtrsim$100K

Avik Halder, Anthony Liang, and Vitaly V. Kresin[*]

Department of Physics and Astronomy,
University of Southern California,
Los Angeles, CA 90089-0484, USA

*e-mail: kresin@usc.edu

A unique property of size-resolved metal nanocluster particles is their "superatom"-like electronic shell structure. The shell levels are highly degenerate, and it has been predicted that this can enable exceptionally strong superconducting-type electron pair correlations in certain clusters composed of just tens to hundreds of atoms. Here we report on the observation of a possible spectroscopic signature of such an effect. A bulge-like feature appears in the photoionization yield curve of a free cold aluminum cluster and shows a rapid rise as the temperature approaches ≈100 K. This is an unusual effect, not previously reported for clusters. Its characteristics are consistent with an increase in the effective density of states accompanying a pairing transition, which suggests a high-temperature superconducting state with $T_c \gtrsim 100$ K. Our results highlight the promise of metal nanoclusters as high-$T_c$ building blocks for materials and networks.

Size-resolved metal nanocluster particles display a number of distinctive quantum size effects, one of the most remarkable being their "superatom"-like electronic shell structure[1,2]. It has been predicted that as a result of the shell levels' high degeneracy, nanoclusters composed of just tens to hundreds of atoms can display pair correlations associated with the superconducting state[3–6]. For certain sizes and materials the pairing may become exceptionally strong. Such effects should have spectroscopic manifestations. We have observed a novel spectral feature which appears in the photoionization curve of some free cold aluminum clusters and rapidly strengthens as the temperature approaches ≈100 K from above. As will be shown below, this behavior reflects an increase in the effective density of states and its temperature dependence. It is consistent with a pairing transition and suggests a high-temperature superconducting state with $T_c \gtrsim 100$ K.

The electronic structure of many free metallic clusters possesses a remarkable feature: the size-quantized states of their delocalized electrons organize into a sharply pronounced "shell structure": multiplets of highly degenerate levels. This is analogous to the principle underlying the atomic periodic table (hence the moniker "superatoms"[7]) and the stability pattern of nuclei. For a spherical (closed-shell, or "magic") cluster the level degeneracy is approximately $2(2L+1)$, where $L$ is the shell angular momentum quantum number. If the topmost shell is not fully occupied, the cluster can undergo a Jahn-Teller deformation away from spherical symmetry.

Qualitatively, the shell degeneracy in a cluster can be viewed as a sharp peak in the density of states (DOS) near the Fermi level, akin to a Van Hove singularity. This amplifies the pairing coupling constant, which is proportional to the DOS, and greatly enhances the gap parameter $\Delta$ and the critical temperature. In some (but by no means all) cluster sizes, a concurrence of large $\Delta$ and sufficiently small[8-10] shell spacing $\delta\varepsilon$ creates a situation very favorable for high-$T_c$ pairing.

The formation of Cooper pairs in this situation has analogy to that in atomic nuclei[11,12]: in both cases the pairs are composed of fermions with opposite projections of orbital and spin angular momenta.

The prospect of high-$T_c$ superconductivity in nanoclusters was remarked upon already by such authors as J. Friedel[13], W. D. Knight[14] and B. Mottelson[15] but the detailed theoretical analysis and its prediction of great strengthening relative to bulk samples appeared more recently, as cited above. Rigorous treatment employs the general formalism appropriate for strong coupling and takes into account the discrete nature of the spectrum and the conservation of particle number, as appropriate for a finite Fermi system. It also affirms that fluctuations of the order parameter will broaden but not destroy the pairing transition: thanks to the large values of $T_c$ and the gap, the coherence length becomes quite small (comparable to the cluster size), hence in this respect the system is not zero-dimensional. Overall, then, the theoretical prediction of high-temperature superconducting pairing in individual size-selected clusters with shell structure rests on solid foundation and calls for experimental verification.

The experimental situation with metal nanocluster superconductivity is rather scarce. Peculiar odd-even effects in the electric susceptibilities of cold ($\approx$20 K) niobium clusters have been hypothesized[16] to result from pairing. These interesting effects merit further exploration, but Nb clusters do not exhibit electronic shell ordering and hence do not fall into the family of potential high-$T_c$ nanoclusters which are at the focus of this paper. Ref. 17 observed small heat capacity jumps at $T\approx$200 K for a pair of free aluminum cluster ions and suggested that they may be consistent with pairing, however the data statistics were very limited. Tunneling measurements[18] at $T$=1.4 K revealed gap variations in surface-grown tin nanoparticles of 1-30 nm height, illustrating that pairing in systems with a discrete energy spectrum can be very sensitive to the number of electronic energy levels falling within the pairing window. Particles studied in this reference did not possess the symmetry required for shell structure with its angular momentum degeneracy, hence they displayed gap enhancement of up to 60%, rather than the orders-of-magnitude increase predicted for free clusters.

**Results**

Although many standard techniques of detecting superconductivity in the bulk cannot be applied to nanoclusters flying in a beam, pairing correlations should have spectroscopic manifestations. In particular, as will be discussed in more detail below, they modify the effective DOS, and this can be detected via the photoionization technique. To search for the first observation of such an effect, we carried out a series of accurate measurements of the photoionization yield curves of free $Al_n$ clusters for several temperatures in the range of 65 K – 230 K. The choice of aluminum was based on the fact that it is a superconductor in the bulk ($T_c$=1.2K) and at the same time many clusters with $n$>40 are well described by the shell model[19–21]. It was one of the materials suggested as a strong candidate in the context of high-temperature pairing in nanoclusters[3].

The beam of neutral clusters was produced in a magnetron sputtering/aggregation source[22]. Their internal temperatures were controlled by passing the beam through a thermalizing tube attached to the source exit, where collisions with helium gas provided equilibration with the tube's walls. Downstream from the exit, the cluster were ionized by a cross beam from a tunable pulsed laser attenuated to ensure single-photon absorption, and the resulting ions were extracted into a linear time-of-flight mass spectrometer. Further details are provided in the Methods section. Mass spectra were acquired for different laser wavelengths in 1 nm steps, and properly normalized intensities of $Al_n^+$ ions were then plotted as a function of photon energy $E$. Such an ion yield plot $Y(E)$ is analogous to a threshold photoemission current measurement in the case of a bulk surface. The data were collected for cluster size ranges $n$=32-95 and temperatures of 65 K, 90 K, 120 K, 170 K, and 230 K. A single experimental run for each temperature lasted approximately 24 hours, and each such run was repeated 4-5 times.

Most cluster ion yield curves display a fairly monotonic post-threshold rise for all temperatures, as illustrated in Fig. 1a. They can be used to derive the cluster ionization thresholds, as will be presented elsewhere, but do not display any abnormalities. However, and this is the main result of this letter, for just a few sizes ($Al_{37,44,66,68}$) with decreasing temperature there appears a bulge-like feature close to the ionization threshold. The clearest and most prominent example is observed in the photoionization spectrum of the closed-shell "magic" cluster $Al_{66}$ with 198 valence electrons[20,21], as seen in the progression of spectra shown in Fig. 1b. Here we will focus the discussion on this system; other sizes will be discussed in detail elsewhere.

In order to characterize the temperature dependence of the $Al_{66}$ spectral feature, we chose two possible procedures. One is to plot the area under the bulge, see Fig. 2c. The other is to differentiate the yield curve, as shown in Fig. 2a. The large peak in $dY(E)/dE$ stems from the bulge, and Fig. 2b represents the amplitude of the derivative maximum as a function of cluster temperature. The plots in Figs. 2b,c are qualitatively similar and together suggest that there is a transition taking place at $T \gtrsim 100$ K.

This appearance of a bulge in the $Al_{66}$ spectrum associated with a decrease in temperature is the most significant part of the observation. The feature far exceeds anything similar resolved among its neighboring clusters. In addition, while hump-like structures in near-threshold ionization curves have been seen in other nanoclusters with shell structure, such as[23,24] $Cs_n$ and $Cs_nO$, none appeared in closed-shell structures and, most significantly, none were reported to be temperature-dependent. To confirm this, we have measured photoionization yield curves for $Cu_n$ clusters, $n$=24-87, over the same range of temperatures as $Al_n$. And indeed, "magic-number" copper clusters did not show any notable structure near threshold and, as illustrated in Fig. 3, whatever structure was present for several open-shell cluster sizes showed absolutely no significant temperature dependence. Both of these factors are in strong contrast to the aluminum case.

**Discussion**

Thus the effect reported in Figs. 1b and 2 appears to be a new observation, and merits careful attention and exploration. We suggest that it is consistent with being a manifestation of the electron pairing phenomenon, as supported by several factors:

- An actual temperature onset is clearly observed. (The gradual decrease in the intensity of the bulge above the transition is consistent with the behavior of pairing fluctuations expected in a finite system[25].)

- The onset lies within the temperature range matching theoretical predictions for aluminum cluster superconductivity. According to theory, the value of $T_c$ depends strongly on the occupied-to-unoccupied shell spacing (HOMO-LUMO gap) and the value of the material's electron-phonon coupling constant $\lambda$. Calculations in Ref. 3 give $T_c \sim 100$ K for closed-shell Al clusters with a coupling constant $\lambda=0.4$, as in crystalline aluminum, and a shell gap of ~0.1 eV. Photoelectron spectroscopy[21] reveals that the intershell distance in $Al_{66}$ is a bit larger, approximately 0.35 eV (taking the distance between the half-maximum points on the facing slopes of the topmost peaks). However, theoretical studies of $Al_n$ clusters[26] suggest that $n=66$ (and possibly some of the other aforementioned sizes) has amorphous-like structure, and therefore the value of $\lambda$ used also should correspond to amorphous aluminum. The latter has a $T_c$ of 6 K (see, for example, Ref. [27]) which significantly exceeds that for usual crystalline Al, 1.2 K. Correspondingly, the value of $\lambda$ here is much larger[28]. With the use of McMillan relations (see, e.g., Ref. [29]) for $T_c$ and for $\lambda$, one can estimate that for amorphous Al the coupling is indeed quite strong: $\lambda \approx 1$-1.2. Note also that $Al_{66}$ has the most spherical shape among its peers and therefore the most pronounced shell system[21] and the highest level degeneracy, which matches the optimal scenario for a pairing mechanism.

- The effect appears only in a few of the clusters studied. This agrees with the expectation that pairing can take place only in systems with a propitious combination of electronic degeneracy, shell energies, and coupling strength. While it is conceivable that thermal structure fluctuations could themselves somehow cause a bulge in the electronic spectrum, the observed onset temperature lies much below the aluminum clusters' pre-melting and melting points[30] of 300-900 K. It would be unexpected for purely structural fluctuations in several clusters of different sizes to conspire to produce a single similar-looking feature in their electronic spectra. Although such a possibility merits bearing in mind, the above considerations strongly suggest that we are dealing not with a structural but rather with an electronic transition.

How can one relate the observed spectroscopic feature to the structure of the nanocluster's energy spectrum? The photoionization yield as a function of photon energy $E$ is given by $Y(E) \propto \int_{-E}^{\infty} M(\varepsilon)f(\varepsilon)D(\varepsilon)d\varepsilon$, where $\varepsilon$ is the electron energy, $M$ the dipole transition matrix element, $f$ the Fermi-Dirac occupation function, and $D$ the electronic density of states. The energy of the vacuum level is set to zero in this expression. The derivative $dY/dE$ is therefore proportional to $D(\varepsilon)$[23,24], and one can surmise that the peak shown in Fig. 2a is a reflection of the electronic DOS and its change with temperature.

In the superconducting scenario the energy of excitations acquires the form $\tilde{\varepsilon} = \left(\xi^2 + \Delta^2\right)^{1/2}$, where $\xi$ is the electron energy in the normal state referred to the chemical potential $\mu$. As a result, the onset of pairing both compresses the highest-occupied electron shell (as is familiar from the pattern in bulk superconductivity) and pushes it downwards[3] (reflecting the extra pair-breaking energy now required to move an electron into the continuum) towards the lower shells which lie quite closely[21]. The consequence is a rise in the near-threshold DOS, as observed.

(Even in an odd-numbered cluster, such as $Al_{37}$ mentioned above, the same scenario can take place: the near-threshold photoemission curve will derive from one unpaired electron, plus a bulge due to the much larger number of the remaining paired electrons.)

It is important to keep in mind that both the order parameter $\Delta$ and the chemical potential $\mu$ depend on the temperature, the latter due to the requirement of particle conservation in a finite system (see, e.g., ref. 3). The dependence $\Delta(T)$ is especially rapid near $T_c$. This translates into the temperature dependence of the observed spectral feature.

While bulk superconductors don't share the key nanocluster characteristic of a discrete spectrum, it is instructive to inquire whether the superconducting transition also has a measurable influence on the photoelectric effect in these systems. Indeed, refs. 31,32 have observed that near-threshold photoelectron yield in $Bi_2Sr_2CaCu_2O_{8+\delta}$ undergoes marked changes below $T_c$. In particular, below $T_c$ the photoyield spectra acquire new structures (which are also detectable as peaks in the d$Y$/d$E$ derivatives) that can be ascribed to features in the superconducting density of states and therefore are analogous to the Al cluster's "bulge" reported here.

The cuprate data[31,32] also demonstrated that the total amount of photocurrent is strongly altered below $T_c$. At present such an absolute yield measurement is inaccessible to an experiment on free nanoclusters because the fluxes of neutral clusters in the beam are themselves affected by the thermalizing tube temperature. It offers an interesting prospect for a future experiment.

In summary, the spectral feature appearing in the ionization yield spectrum of a cold aluminum nanocluster is indicative of a transition in the electronic density of states. It is a heretofore unobserved phenomenon in cluster spectroscopy. Moreover, it is consistent with the onset of electron pairing at a temperature above 100 K. It will be interesting to extend the exploration of this behavior to other potentially superconducting nanoclusters with quantized electron shell ordering. An attractive prospect is to develop such size-selected nanoclusters into building blocks for Josephson tunneling networks and novel high-$T_c$ materials.

**Methods**

A continuous beam of neutral nanocluster particles is produced by a magnetron/condensation source. Atoms sputtered out of a metal target are entrapped within helium flowing through the source body (a liquid nitrogen cooled tube) and nucleate into clusters. The flow then enters a thermalizing tube of 12 cm length and 1.6 cm inner diameter bolted onto the source. Its temperature was adjusted from 65 K - 230K and the nanoparticles undergo ~$10^5$ collisions with the buffer gas, thereby equilibrating with the tube wall to within at most a few K[33,34]. Downstream from the tube exit they are ionized by nanosecond pulses from a tunable laser system (Ekspla NT342/3/UV). The light fluence is monitored and maintained at ≈500 µJ/cm$^2$ to ensure that the signal lies well within the linear (single-photon) regime. The ionized clusters are then mass separated by a 1.3 m long on-axis linear time-of-flight mass spectrometer and counted by a channeltron detector. The mass spectrum is deconvoluted using multi-peak Gaussian fitting. The yield values are obtained as $Y_n(E)=\bar{I}_n(E)/\phi(E)$, where $\phi$ is the laser fluence and $\bar{I}$ is the normalized $Al_n^+$ counting rate. The normalization corrects for any possible beam intensity drifts by referring the ion rate to that measured at the wavelength of 216 nm after each collection interval. The resulting ionization yield curves are highly reproducible. In particular, the appearance of the "bulges" and their evolution with temperature were confirmed by repeated measurements, as was the absence of detectable temperature-dependent features in the spectra of other clusters.

**Acknowledgement**

This work is supported by the U. S. National Science Foundation under Grant No. DMR–1206334.

**Figures**

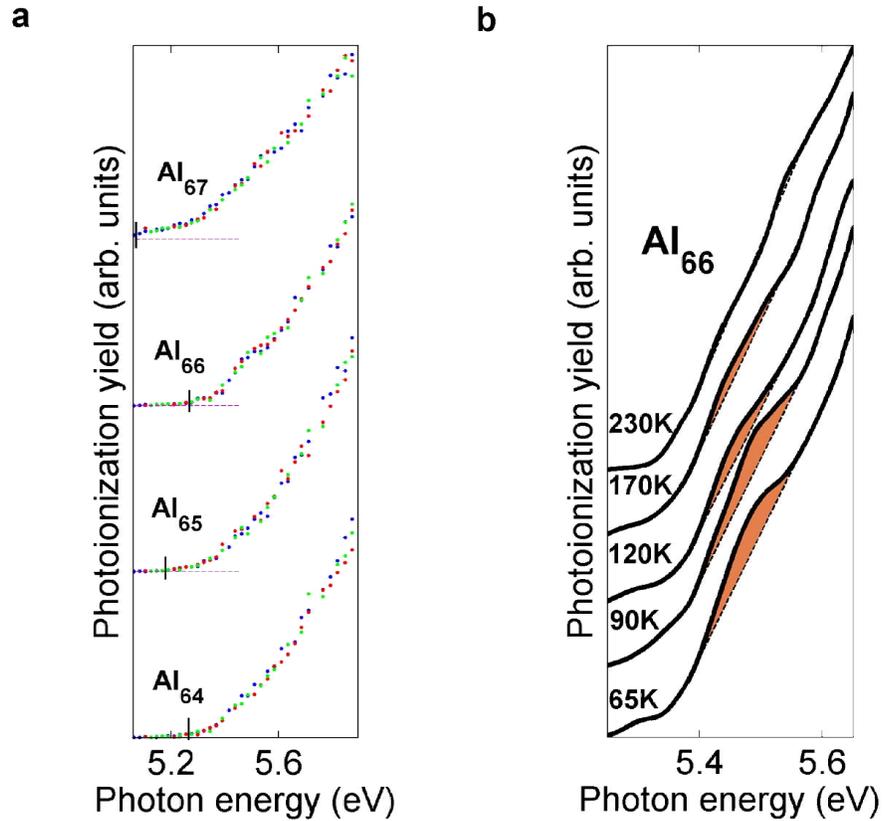

**Figure 1. Photoionization of aluminum nanoclusters.** (a) Yield plots for $Al_n$ ($n=64,\ldots,67$) obtained at $T=65K$. Short vertical bars denote the cluster ionization threshold energies. A strong bulge-like feature appears close to the threshold for $n=66$. The adjacent clusters show no such feature. The sharp drop in the ionization energy from $Al_{66}$ to $Al_{67}$ reflects the fact that the latter is a closed-shell "magic"-size cluster. Different color dots correspond to data duplicated over several experimental runs. In both panels the different yield curves are shown shifted with respect to each other for clarity. (b) Evolution of the $Al_{66}$ spectral feature. Its growth with decreasing temperature can be seen by comparing the thick experimental yield curve (a spline average of the data points, such as shown in the first panel) with the dashed interpolating line.

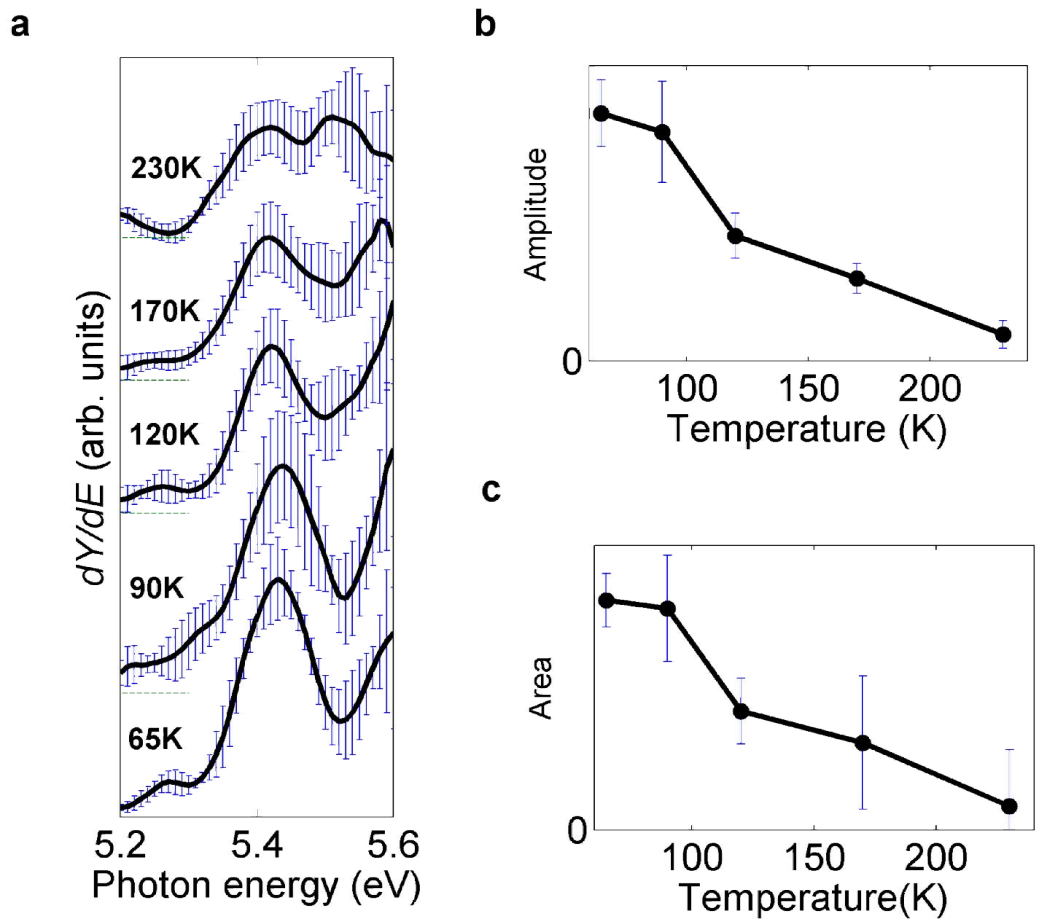

**Figure 2. Temperature dependence of the Al$_{66}$ spectrum and the density of states.**
**(a)** Derivatives of the near-threshold portion of the photoionization yield plots from Fig. 1(b). As discussed in the text, *dY/dE* represents a measure of the electronic density of states. The intensity of the first peak, which derives from the "bulge" in the Al$_{66}$ spectrum, grows with decreasing temperature, implying a rise in the density of states near threshold. The plots are normalized to the amplitude height of the minimum following the derivative peak. **(b)** To quantify the intensity variation of the peak in (a), we plot its amplitude as a function of cluster temperature. **(c)** Another measure of the magnitude of the bulge: its area relative to the dashed straight line in Fig. 1(b). It is noteworthy that the behavior of the plots in panels (b) and (c) matches, both suggesting that a transition takes place as the temperature approaches ≈100 K.

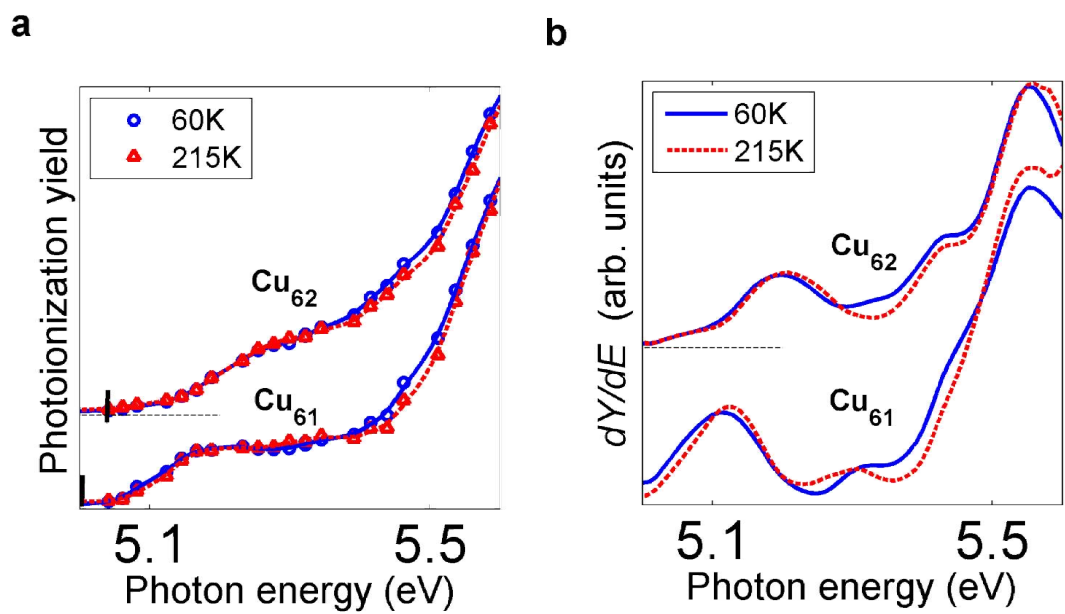

**Figure 3. Photoionization yield curves for copper nanoclusters (a)** and their derivatives **(b)**, illustrated here for two representative sizes, show no temperature dependence similar to that of the detected spectral feature in $Al_{66}$ (Fig. 1). This confirms that the latter case represents a distinctive electronic transition.